\documentstyle[preprint,aps,prb]{revtex}
\def\bbbone{\hat{\mathchoice {\rm 1\mskip-4mu l} {\rm 1\mskip-4mu l}
{\rm 1\mskip-4.5mu l} {\rm 1\mskip-5mu l}}}

\begin{document}

\title{Correlated Wave-Functions and the Absence of Long Range Order in
Numerical Studies of the Hubbard Model} 
\author {M.\ Guerrero, G. Ortiz  and J. \ E. \ Gubernatis}
\address{
Theoretical Division, Los Alamos National Laboratory, Los Alamos, NM 87545}
\date{ \today}

\maketitle
\begin{abstract}
We present a formulation of the Constrained Path Monte Carlo (CPMC) 
method for fermions that uses trial wave-functions that include 
many-body effects. This new formulation allows us to implement a whole 
family of generalized mean-field states as constraints. As an example, 
we calculated superconducting pairing correlation functions for the 
two-dimensional repulsive Hubbard model using a BCS trial state as 
the constraint. We compared the results with the case where a 
free-electron trial wave-function is used. We found that the correlation 
functions are {\it independent} of which state is used 
as the constraint, which reaffirms the results previously found by 
Zhang et. al \cite{shiwei2} regarding the suppression of long range 
pairing correlations as the system size increases.
\vspace{0.5cm}

\end{abstract}


\narrowtext

\section{Introduction}
Since the discovery of high temperature superconductivity, an enormous
effort has been devoted to the theoretical study of two-dimensional electronic
models. This effort is driven by the belief that the mechanism for
superconductivity lies within the CuO$_2$ planes common to these
materials and is dominantly electronic in origin.  The two-dimensional
repulsive Hubbard model  has attracted the
most attention as the simplest  effective model possibly embodying 
the key electronic phenomena at low energies. Numerous works on 
this model have reproduced qualitatively
the observed magnetic properties of the cuprates in the normal 
state. \cite{elbio} However, the search for superconductivity in the 
Hubbard model, although intensive and extensive, has yielded few 
positive indicators. \cite{elbio}

Most of the present knowledge on the phase diagram of the two-dimensional 
repulsive  Hubbard model has been obtained by combination of theorems
and  numerical studies of finite size clusters. The numerical studies 
used Lanczos, Variational Monte Carlo, and zero or finite
temperature quantum Monte Carlo techniques. In a superconducting phase, one
expects the superconducting pairing correlation functions to exhibit off-diagonal long
range order (ODLRO), which is an indication of the Meissner effect. \cite{sewell} 
With this in 
mind, a number of investigators have
calculated pairing correlation functions in various symmetry
channels. However, most calculations were limited to high
temperatures and small system sizes. In the case of Monte Carlo
studies these limitations were imposed by the fermion sign
problem which causes the variances of computed quantities and hence the 
computing time to grow exponentially with the increase in system sizes.

Recently, a new zero temperature quantum Monte Carlo method, the
Constrained Path Monte Carlo (CPMC), was developed that overcomes the major
limitations of the sign problem. \cite{shiwei1} This method allows 
the calculation of pairing correlation functions at zero temperature 
without the exponential
increase in computer time with system size. Using this method, Zhang
et al. \cite{shiwei2} calculated $d_{x^2-y^2}$-wave and extended 
s-wave pairing correlation functions
versus distance in the ground state for lattices up to 
$16 \times 16$.
They found that the $d_{x^2-y^2}$-wave correlations are stronger than extended
s-wave correlations. However, as the system size or the interaction
strength was increased, the magnitude of the long-range part of both
correlation functions vanished.  

Although the findings of Zhang et al. \cite{shiwei2} provide 
evidence for the absence of ODLRO in the two-dimensional Hubbard model,
the CPMC method is approximate and has a systematic error which is
difficult to gauge. The systematic error is associated with the wave-function
used to constrain the Markov chains produced by the Monte Carlo
procedure. 
More specifically, in the CPMC method the ground state wave-function is 
represented by an ensemble of
Slater determinants. As these determinants evolve in imaginary time,
the ones with a negative overlap with a constraining wave-function are
discarded. This procedure eliminates the sign problem but introduces
an approximation that depends on the quality of the constraining 
wave-function. Zhang et al. \cite{shiwei2} used free-electron and unrestricted
Hartree-Fock wave-functions. More
sophisticated choices of wave-functions, particularly ones exhibiting
strongly correlated electron effects, are typically difficult to
implement, because of the increasing number of Slater determinants
needed and the consequent increase in computing time.  

In this work, 
we extended the formulation of the CPMC method in
a way that allows the use of a wide variety of trial wave-functions with
only a small increase in computing time. As
an illustration, we calculated the superconducting pairing correlation functions of
the two-dimensional repulsive Hubbard model in the $d_{x^2-y^2}$-wave channel
using as a constrain a BCS wave-function that has superconducting ODLRO. We 
found that the resulting correlation functions are the same as those
obtained using the free-electron and Hartree-Fock constraining wave-functions. 
This reaffirms the results by Zhang et al. \cite{shiwei2} regarding the 
vanishing of long range pairing correlations as the system size 
increases. 

The article is organized as follows: in section \ref{METHOD} we briefly 
describe the CPMC technique emphasizing aspects of the new formulation. 
In section \ref{RESULTS} we define the Hamiltonian and pairing correlation 
functions and present our results. In section \ref{CONCLU} we discuss our 
conclusions.

\section{\label{METHOD} Method}

In this section we summarize the main features of the CPMC method. For a more 
detailed description of the method see Ref.\onlinecite{shiwei1}.
In the CPMC method, the ground-state wave-function
$|\Psi_0\rangle$ is projected in imaginary time $\tau$ from a known initial 
wave-function $ | \Psi(\tau=0) \rangle = |\Psi_T\rangle$ by a branching 
random walk in an over-complete space
of Slater determinants $|\phi\rangle$,

\begin{eqnarray}
| \phi \rangle = \prod^{N_\sigma}_{i, \sigma} \phi^\dagger_{i \sigma} | 0 
 \rangle \, \, \, \, ; \, \, \, \, \phi^\dagger_{i \sigma} = \sum^N_{j=1} 
c^\dagger_{j \sigma} \Phi^\sigma_{ji} \ ,
\end{eqnarray}
where $c^\dagger_{j \sigma}$ creates and electron in orbital $j$ 
with spin $\sigma$ ($n_{j \sigma} = c^\dagger_{j \sigma} 
c^{\;}_{j \sigma}$), and 
\begin{eqnarray}
\langle \phi | \phi^\prime \rangle \ne \delta_{\phi \phi^\prime}
\end{eqnarray}
with $N$ the number of available single-particle states 
(for the Hubbard model corresponds to the total number of lattice 
sites) and $N_\sigma$ the number of particles 
with spin $\sigma$. The total number of electrons is given by 
$N_e = N_\uparrow + N_\downarrow$.

The projection corresponds to finding  the
ground-state from the long-time solution of the imaginary-time
representation of Schr\"odinger's equation specified by a Hamiltonian 
$\hat{H}$
\begin{equation}
 \frac{\partial|\Psi\rangle}{\partial\tau} = -
(\hat{H}-E_0\bbbone)|\Psi\rangle
\end{equation}
with $E_0$ the ground-state energy ($\hbar$ is set to $1$). 

Provided 
$N_0=\langle\Psi_0|\Psi(0)\rangle\not=0$ and $\hat{H}$ being time-independent, 
 the formal solution
\begin{equation}
  |\Psi(\tau)\rangle = e^{-\tau (\hat{H}-E_0\bbbone)}|\Psi(0)\rangle
\end{equation}
has the property
\begin{equation}
  \lim_{\tau \rightarrow \infty} |\Psi(\tau)\rangle 
     = N_0|\Psi_0\rangle
\end{equation}
On the computer this large $\tau$ limit is accomplished by breaking up
 $\tau$ in small time-steps $\Delta \tau$ and iterating the equation
\begin{equation}
  |\Psi^{n+1} \rangle =
e^{-\Delta\tau(\hat{H}-E_T\bbbone)}|\Psi^n\rangle \label{IT}
\end{equation}
where $E_T$ is a guess at the ground-state energy $E_0$ and 
$\Delta \tau N_s = \tau$ with $N_s$ the number of imaginary time-steps.
As $\tau\rightarrow \infty$, the iteration becomes 
stationary, i.e. $\partial|\Psi\rangle/\partial\tau =0$,  and if $E_T$ is
 adjusted to equal $E_0$, then $|\Psi(\tau \rightarrow \infty) \rangle = N_0|\Psi_0\rangle $.

The propagation in imaginary time is done in the following way: 
in the space of Slater determinants, we write $|\Psi_0\rangle =
\sum_\phi \chi(\phi) |\phi\rangle$ and choose $\chi(\phi)>0$. By being
positive, the function $\chi(\phi)$ describes the distribution of Slater
determinants representing the ground state. The Monte Carlo process
samples from this distribution. This process is implemented by the
application of a Trotter decomposition  and a Hubbard-Stratonovich
transformation to the iterative equation ~(\ref{IT})  and converting it into
\begin{equation}
 |\Psi^{n+1}\rangle = \int d{\bf x}\, P({\bf x}) B({\bf x})|\Psi^n\rangle
\label{INTEGRAL}
\end{equation}
where ${\bf x}$ is a multi-dimensional random variable distributed according 
to $P({\bf x})$ and $B({\bf x})$ is an operator
approximating $e^{-\Delta\tau \hat{H}}$ for a given value of the random
variable, whose general structure is a product of exponentials of operators 
quadratic in $c$ and $c^\dagger$. For each time 
step $\Delta \tau$, $B({\bf x})$ has the property 
of transforming one Slater determinant into another.  The Monte Carlo 
method evaluates
the multi-dimensional integral ~(\ref{INTEGRAL}) by using an ensemble of
random walkers represented by  Slater determinants
$|\phi\rangle$. For each walker, it samples  ${\bf x}$ from $P({\bf
x})$ and then obtains the new Slater determinant by multiplying
\begin{equation}
 |\phi^{n+1}\rangle = B({\bf x})|\phi^n\rangle
\end{equation}
Once the Monte Carlo procedure converges, the ensemble of
$|\phi\rangle$ represents
$|\Psi_0\rangle$ in the sense that their distribution is a Monte Carlo
sampling of $\chi(\phi)$. In this sense, the CPMC approach is a sort of 
stochastic configuration interaction method.

To specify the ground-state wave-function completely, only
determinants satisfying $\langle\Psi_0|\phi\rangle>0$ are needed
because $|\Psi_0\rangle$ resides in either of two degenerate halves of
the Slater determinantal space (in general, a manifold of dimension 
$N_e(N-N_e)$), separated by a nodal hypersurface ${\cal N}$
defined by $\langle\Psi_0|\phi\rangle=0$.  The sign problem
occurs because walkers can cross ${\cal N}$ as their orbitals evolve
continuously in the random walk. Asymptotically in $\tau$ they populate the two
halves equally, leading to an ensemble that tends to have zero overlap with
$|\Psi_0\rangle$.  If ${\cal N}$ were known, one would simply constrain
the random walk to one half of the space and obtain an exact solution
of Schr\"odinger's equation.  In the CPMC method,
without {\it a priori\/} knowledge of ${\cal N}$, we use a
constraining wave-function, which we usually take to be the 
trial wave-function $|\Psi_T\rangle$, and require the Slater determinants
to satisfy 
$\langle\Psi_T|\phi\rangle>0$.  Thus, the quality of the calculation clearly
depends on $|\Psi_T\rangle$. In the past only free-electron or
Hartree-Fock wave-functions were implemented, mainly due to their
simplicity and the novelty of the method.  However, it is desirable to
use more sophisticated wave-functions that include many-body effects. 
For example, to study superconductivity it is interesting to 
implement trial wave-functions that exhibit ODLRO, like a BCS wave-function.

Our goal is to use trial wave-functions of the type (i. e., a Bogoliugov
transformation of the vacuum $|0 \rangle$, $ \langle 0|0 \rangle =1$)
\begin{eqnarray}
| \Psi_T\rangle =\prod_k(u_k + v_kc^\dagger_{k \uparrow}c^\dagger_{-k \downarrow})|0\rangle
\label{PHI}
\end{eqnarray}
where the product includes all values of momentum $k = (k_x,k_y)$ in the first 
Brillouin zone and $|u_k|^2 + |v_k|^2 =1$ to ensure normalization
($\langle \Psi_T | \Psi_T \rangle = 1$).
Other than satisfying the normalization condition, the parameters $u_k$ 
and $v_k$ can be chosen arbitrarily. 

Equation (\ref{PHI}) represents a wave-function that does not have a
fixed particle number $N_e$. To represent a fixed electron number,
$|\Psi_T\rangle$ needs to be projected onto that particular subspace.
The resulting wave-function is a linear combination of a large number
of Slater determinants \cite{bouchaud} (large in the sense that the number  
grows very rapidly
with system size and particle number to the point where it becomes
impractical to use).  Alternatively, one can work in an extended space with
different electron numbers. To do that, we follow Yokoyama and 
Shiba \cite{yoko} 
and perform a particle-hole transformation on one of the spin species:

\begin{equation}
 \left\{  \begin{array}{rcl}
 d_k^{\;} &=& c^\dagger_{-k \downarrow} \\
 c^\dagger_k &=& c^\dagger_{k \uparrow}
\label{PHTRANS}
          \end{array} 
\right. 
\end{equation}
Using this transformation and noting that the new vacuum 
$|\widetilde{0}\rangle$ is related to the old one by
\begin{eqnarray}
|0\rangle = \prod_k d^\dagger_k |\widetilde{0}\rangle
\end{eqnarray}
we can rewrite $|\Psi_T\rangle$ in terms of the new $c$ and $d$ operators:
\begin{eqnarray}
| \Psi_T\rangle =\prod_k(u_k d^\dagger_k + v_kc^\dagger_k)|
\widetilde{0}\rangle
\label{PHI2}
\end{eqnarray}
so that $| \Psi_T \rangle$ is represented by a single Slater determinant.
Since we are interested in projecting out the ground state with a fixed
electron number, we have
to use the propagator $e^{- \tau (\hat{H} - E_0 \bbbone - \mu 
\hat{N}_e)} = \hat{\cal U}(\tau)$ and choose $\mu$, the chemical 
potential,  to select 
the desired number of electrons $N_e =  \langle \Psi_0 | 
\hat{N}_e | \Psi_0 \rangle/
\langle \Psi_0 | \Psi_0 \rangle $ ($\hat{N}_e = \sum_{j \sigma} 
n_{j \sigma}$). At the end of the
projection the ground state wave-function will have a fixed number of
electrons given by the choice of $\mu$.

The changes in the CPMC method necessary to use the BCS form of a
correlated wave-function are minor. Instead of 
matrices $\Phi^\sigma$ for up and down spin of sizes $N \times N_\sigma$
to represent the random walkers,
they, as well as the trial wave-function $| \Psi_T \rangle$, are now represented by a
single matrix of size $2N \times N$. 
The increase in computation time caused by the increase in
 the size of the matrices depends on the 
system size and the number of particles. A rough estimate gives the 
increase as the factor  $3N/N_e$. 
For example, for a $6 \times 6$ system with  $N_e=26$ 
this is  $4 = 2.89 N/N_e$. The 
closer we get to 
half-filling ($N_e=N$) the smaller the increase. In general,
for the filling fractions studied here, the increase in computer time is
of the order of 4.

\section{\label{RESULTS} Calculation and Results}

The Hamiltonian is the usual Hubbard Hamiltonian on a square lattice with 
periodic boundary conditions:
\begin{eqnarray}
\label{HAMILTON}
\hat{H}= -t\sum_{<ij>,\sigma} (c^\dagger_{i,\sigma}c_{j,\sigma}^{\;} + 
c^\dagger_{j,\sigma}c^{\;}_{i,\sigma}) + U \sum_{i} n_{i \uparrow}n_{i \downarrow}
\end{eqnarray}
where $t$ is the nearest neighbor hopping matrix element and  $U$ is the on-site Coulomb 
repulsion. We set $t=1$ so that all energies are measured in units of $t$.
In terms of the operators $c$ and $d$ defined by the transformation 
~(\ref{PHTRANS}) the Hamiltonian has the form
\begin{eqnarray}
\label{HAMILTON2}
\hat{H}= -t\sum_{<ij>} (c^\dagger_ic^{\;}_j + c^\dagger_jc^{\;}_i - 
d^\dagger_id^{\;}_j - d^\dagger_jd^{\;}_i) + 
     U \sum_{i} n^c_i(1-n^d_i)
\end{eqnarray}
where $n^c_i$ $(n^d_i)$ denotes the occupation in the $c$ $(d)$ orbital.
This transformed Hamiltonian corresponds to a two-band spinless fermion model.

We computed the ground-state energy and the superconducting pairing correlation functions
in the $d_{x^2-y^2}$-wave channel using the following
definitions:
\begin{equation}
P_d(\vec{R}) = \langle \Delta_d^\dagger(\vec{R}) \Delta_d(0) \rangle
\end{equation}
where the pair field operator is
\begin{eqnarray}
& \Delta_d(\vec{R}) =  \sum\limits_{\vec{\delta}} f_d(\vec{\delta})  
[c_{\vec{R}  \uparrow}c_{\vec{R}+\vec{\delta} \downarrow} -
c_{\vec{R}\downarrow}c_{\vec{R}+\vec{\delta}   \uparrow} ]
\end{eqnarray}
with $\vec{\delta} = \pm \hat{x}, \pm \hat{y}$ ,  $f_{d}(\pm \hat{x}) = 1 $ 
and $f_{d}(\pm \hat{y}) = -1 $ . $\vec{R}$ denotes the position in the lattice
in units of the lattice constant which is taken to be unity.

We used  trial wave-functions of the form ~(\ref{PHI}) with $u_k$ and $v_k$ 
given by the BCS relation

\begin{eqnarray}
\label{BCS}
\frac{v_k}{u_k} = \frac{\Delta_k}{\epsilon_k - \mu + \sqrt{(\epsilon_k - \mu)^2 + |\Delta_k|^2}}
\end{eqnarray}
where  $\epsilon_k$ is a single particle energy and 
$\Delta_k$ is the gap, $\Delta_k = \Delta f(k)$. $\Delta$ is a variational
c-number and $f(k)$ represents
the symmetry of the pairing which we choose to be $d_{x^2-y^2}$, 
$f(k)=\cos(k_x)-\cos(k_y)$.

We concentrated in the $d_{x^2-y^2}$-wave channel in part because the 
existence of ODLRO in the extended s-wave channel is conditioned upon the
existence of ODLRO in the isotropic s-wave channel. \cite{sczhang} 
Since the possibility of pairing in the isotropic s-wave channel is 
highly unlikely for the repulsive Hubbard model, so is the chance of
pairing in the extended s-wave channel. Moreover, these statements 
have been verified numerically by us and by Zhang et al. \cite{shiwei2}
Also, it has been increasingly established experimentally that the order
parameter in the superconducting cuprates has $d_{x^2-y^2}$-wave symmetry.

We used two different trial wave-functions: one with $\Delta=0.5$, 
which corresponds to a BCS superconducting state, 
and the other one with  $\Delta = 0$, which corresponds to the free-electron 
case. In both cases we choose the parameter $\mu$ in the BCS wave-function
so that $\langle \Psi_T | \hat{N_e} | \Psi_T \rangle = N_e$ where
$N_e$ is the number of electrons we are interested in.
While the free-electron wave-function has a fixed number of electrons
$(\sigma_{N_e} = 
\sqrt{\langle \hat{N_e}^2 \rangle - \langle \hat{N_e} \rangle^2} = 0)$, 
the BCS wave-function with $\Delta \neq 0$ has components with different 
electron numbers so that $\sigma_{N_e} \neq 0 $. 
It is important to notice that in general the parameter $\mu$ in the BCS 
wave-function is different than the one used in the propagator 
$\hat{\cal U}(\tau)$. The latter one is set so that at the end of 
the propagation the ground state has the desired number of electrons $N_e$.

To illustrate the difference between these two wave-functions, in Fig. 1 
we plot the variational value of the $d_{x^2-y^2}$-wave correlation 
functions versus distance, that is 
$ \langle \Psi_T | \Delta_d^\dagger(\vec{R}) \Delta_d(0) | \Psi_T 
\rangle $, for the two trial wave-functions 
in a $10 \times 10$
system with $U=4$ and $N_e = 82$, so that the filling fraction is 
$n_e = N_e/N = 0.82$. This filling corresponds to a closed shell case, 
that is, the free-electron ground state is non-degenerate.
 In the free-electron 
case the correlations die out rapidly with distance, while in the BCS case
the existence of ODLRO is evident in the sense that for long distances,
the correlation functions approach a finite value given by the square of
the superconducting order parameter $\Delta^{SC}$:
\begin{equation}
\Delta^{SC} = \frac{4}{N} \sum_k f(k) u_k v_k = \frac{4}{N} \sum_k f(k) 
\frac{\Delta_k}{\sqrt{(\epsilon_k - \mu)^2 + \Delta_k^2}}
\end{equation}
The overlap between the two normalized trial wave-functions is
$\langle \Psi_T(\Delta=0) | \Psi_T(\Delta=0.5) \rangle = 0.0076$,
so the two wave-functions are close to being orthogonal.

The variational energy $E_v = \langle \Psi_T | \hat{H} | \Psi_T \rangle$ is 
much larger for the BCS trial 
wave-function than for the free-electron trial wave-function.
In general we find that the variational energy increases monotonically with 
the parameter $\Delta$ of the BCS wave-function, as it is shown in Fig. 2 
for a $10 \times 10$ system with $U=4$ and
$\langle \hat{N_e} \rangle = 82$. 
This variation contrasts previous results obtained with the 
Variational Monte Carlo method, which found 
that a non-zero value of $\Delta$ minimizes the variational 
energy. \cite{yoko}$^,$\cite{gross}$^,$\cite{nakanishi} 
However, in these cases, a Gutzwiller factor was included in the wave-function
that projected out totally or partially the states with double occupancy.
It seems that the inclusion of this factor is crucial to obtain a minimum
of the variational energy at a finite value of $\Delta$. At present, our 
formulation does not allow the use of trial wave-functions that are non-Fock
states such as the Guztwiller wave-function:
\begin{equation}
| \Psi_G \rangle = \prod_i (\bbbone-g \ n_{i \uparrow}n_{i \downarrow}) |\Psi_{FOCK} \rangle
\end{equation}
with $g$  a variational parameter that determines the average number
of doubly occupied sites. (When $g=1$,  double occupation is completely 
suppressed.) Even though such wave-functions are not implemented, since we
are doing a projection in imaginary time onto the ground state of the system,
 it is not crucial to improve the variational energy of our trial state.

In the large $U$ limit, the Hubbard model can be mapped onto the $t-J$ model
. This strong coupling limit was used in Refs. \onlinecite{gross} and 
\onlinecite{yoko}  
to calculate the energy, making a comparison with our work difficult.
However, we can do a comparison 
with Ref. \onlinecite{nakanishi} since they used the Hubbard  
Hamiltonian to calculate the energy. In their Fig. 1 they report
the variational energy per site as a function of $\Delta$ for a $6 \times 6$
system with $U=8$, $32$ electrons, periodic boundary conditions in 
the $x$ direction and anti-periodic 
in the $y$  direction. From their figure, the minimum value 
for the energy per site is -0.65523 and corresponds to a value of 
$\Delta = 0.1$. The variational energy per site that we obtain for 
the same system but with periodic boundary conditions in both directions is 
0.02726. The difference can likely be accounted for by the fact that we did 
not project our wave-function onto a fixed particle number and second, we 
did not use
a Gutzwiller factor. However, the ground state energy per site 
calculated with the CPMC method 
is $-0.7272 \pm 0.0005$, which is considerably lower than their value. 

As a check of our algorithm we compared the correlation functions 
and ground-state energy given by 
the CPMC method using the free-electron trial wave-function with
results by Zhang et al., \cite{shiwei3} who used the original 
formulation of the CPMC, for 
a $6 \times 6$ system with $U=4$ and $N_e=26$ 
and  an $8 \times 8$ system with $U=8$ and $N_e=50$. We found excellent 
agreement with their  results. 

In Fig. 3 we plot the resulting correlations functions given by the CPMC
calculation  with the two trial wave-functions used in Fig. 1, for 
10 $\times$ 10 with $U = 4$. 
It is clear that the results are essentially the {\it same} no matter what trial 
wave-function is used. The long distance magnitude of the correlation 
functions is 
very small, smaller than the free-electron case.

Similar calculations to the ones presented in Fig. 3 were done for 
$8 \times 8$ and $6 \times 6$ 
systems with $U=4,6$ and $8$ and dopings corresponding to closed shells 
cases. The results are consistently the same: 
the correlation functions are the same no matter what trial wave-function
is used. The ground-state energy, however, is always larger when the BCS 
wave-function is used. The difference between the two ground-state energies
is larger for larger $U$. When the BCS wave-function is used, we find that
there are more nodal crossings; that is, more walkers  are
discarded because their overlap with  the trial wave-function is negative.
We believe this is why the energy is higher in the case of the 
BCS wave-function.

We did not use systems larger than $10 \times 10$ in part because as
system size increases, it becomes more difficult to select $\mu$ in the
propagator to get the desired number of electrons. This is because the energy
levels are getting closer in larger systems. Also, we found that the 
correlation functions are the same no matter which trial wave-function is 
used for $6 \times 6$, $8 \times 8$ and  $10 \times 10$ systems. This
evidence is enough to conclude that the correlation functions are independent
of which trial wave-functions is used.

\section{\label{CONCLU} Conclusions}

We presented a formulation of the CPMC method that uses
trial wave-functions that include correlation effects and have components
of different electron numbers. Instead of projecting it onto a subspace 
with fixed number of electrons, we used a
particle-hole transformation in one of the spin species to write
such trial wave-functions as only one Slater determinant.

Because of the increase in the size of the matrices used, 
this formulation involves a small increase in computing time  
compared to the original formulation. The increase in CPU
time is roughly $3N/N_e$. For the dopings considered in this work 
it comes to a factor of approximately $4$.

This new formulation is very general and allows the implementation of a whole 
family of mean-field wave-functions. Following Bach, Lieb and Solovej 
\cite{bach} we call this class of functions generalized Hartree-Fock states, 
i. e., states that are ground states of some quadratic mean-field Hamiltonian 
in Fock space which do not necessarily conserve particle number. Possible 
examples include spin-density wave, charge-density wave and 
superconductivity.

As an illustration, and because of its importance in high temperature 
superconductivity, we used a BCS trial wave-function with $d_{x^2-y^2}$-wave
symmetry to calculate the superconducting pairing correlation 
functions in the ground state for the two-dimensional repulsive 
Hubbard model. We compared this result with the one using the 
free-electron trial wave-function. We studied $6 \times 6$, $8 \times 
8$, and  $10 \times 10$ systems for different values of $U$ and 
dopings and found that the results for the correlation functions are 
independent of which trial wave-function is used for the constraint. 

Most of the calculations presented in this work correspond to closed 
shell cases, that is, electron fillings with a non-degenerate free-electron
ground state. To check the consistency of our results we also studied
some open shell cases like a $6 \times 6$ system with $32$ electrons
($n_e=0.89$), $U=8$  and 
periodic boundary conditions. We used three different trial wave functions:
one free-electron wave function with a fixed number of electrons, another
free-electron wave function but with some paired electrons in the Fermi surface
and a BCS wave-function with $\Delta = 0.1$. The CPMC result is consistent
with those of the closed shell cases: the superconducting pairing 
correlation functions, which vanish for large distances,  
are independent of the trial wave-function used. 
Technically, the open shell case is more difficult 
because in general the free-electron trial wave-functions do not 
have translational invariance. For this reason, 
one finds different values of the correlation functions for the same distance
$|\vec{R}|$  but different directions in the lattice. To overcome this 
problem we averaged the correlation functions for a given  $|\vec{R}|$
 over all possible  directions in the lattice. This procedure is also 
used for the closed shell cases but is more relevant in the open shell 
case where the differences are caused by  a broken symmetry introduced by the
trial wave-function as oppossed to  small statistical fluctuations due to 
the Monte Carlo process.

These results reaffirm the previous ones by Zhang et al. 
\cite{shiwei2} implying the absence of ODLRO in the 
$d_{x^2-y^2}$-wave channel of the two-dimensional repulsive Hubbard 
model. We do not dismiss the possibility of ODLRO existing in some 
exotic channel or for some combination of quasiparticle operators instead
of the bare ones. \cite{elbio2} This work has only 
investigated the channels commonly 
studied. Although it is not rigorously proven that the absence of 
ODLRO implies no Meissner effect and consequently no superconductivity, 
it is reasonable to think that a model without apparent ODLRO is 
inappropriate as a model of the superconducting phase for the 
high temperature superconducting materials. 

The lack of clear numerical evidence of $d_{x^2-y^2}$-wave superconductivity 
upon doping and the abundance of clear numerical evidence of
antiferromagnetism at half filling makes it hard to see how a theory,
like the SO(5) phenomenology, can apply to the Hubbard model as some
have recently suggested. \cite{hanke} This phenomenology requires 
the antiferromagnetic long range order at half-filling to transform
into $d_{x^2-y^2}$-wave superconducting long range order in the 
doped states. If the low lying
excited states have approximate SO(5) symmetry, why then does the strong
antiferromagnetic state transform into something that is so
hard to find? The two-dimensional repulsive Hubbard model seems to be 
an inappropriate candidate for the SO(5) phenomenology.

\section{Acknowledgments}
We are thankfull to S. Trugman for the critical reading of the manuscript.
The C++ program used
for this work incorporated the {\it MatrixRef} matrix classes written by S.R.
White, available at http://hedrock.ps.uci.edu. This work was supported by 
the Department of Energy. Some of the calculations were performed on 
the computers at NERSC.

\begin{figure}

\caption{
Variational value of the pairing correlations versus distance $|\vec{R}|$ for 
two different 
trial wave-functions in a $10 \times 10$ system. Parameters are $U=4$
and filling fraction $n_e = 0.82$. The BCS wave-function exhibits ODLRO. 
}
\label{fig1}

\vspace*{0.9cm}

\caption{BCS variational energy per site as a function of $\Delta$ for 
the same system 
as in Fig. 1. The energy increases monotonically with $\Delta$. The inset shows
smaller values of  $\Delta$ where Ref. 9 finds a minimum.
}
\label{fig2}
\vspace*{0.9cm}

\caption{Pairing correlation functions in the $d_{x^2-y^2}$-wave channel 
given by the CPMC method for  same system as in Fig. 1. The inset shows 
the long range part in detail. The results are the same for the two
different trial wave-functions: the correlations decay quickly with distance. 
Errors bars are smaller than the size of the symbols.
}
\label{fig3}
\vspace*{0.9cm}

\end{figure}

\end{document}